\documentclass[utf8]{frontiersSCNS} 
\usepackage[utf8]{inputenc}
\usepackage{url,hyperref,lineno,microtype,subcaption}
\usepackage[onehalfspacing]{setspace}
\def\keyFont{\fontsize{8}{11}\helveticabold}

\def\firstAuthorLast{Plachy {et~al.}} 
\def\Authors{Emese Plachy\,$^{1,2,3}$, R\'obert Szab\'o,$^{1,2,3*}$}
\def\Address{$^{1}$Konkoly Observatory, Research Centre for Astronomy and Earth Sciences, Konkoly Thege Mikl\'os \'ut 15-17, H-1121 Budapest, Hungary \\
$^{2}$MTA CSFK Lend\"ulet Near-Field Cosmology Research Group \\
$^{3}$ELTE Eötvös Loránd University, Institute of Physics, Budapest, Hungary
}
\def\corrAuthor{R\'obert Szab\'o}

\def\corrEmail{szabo.robert@csfk.mta.hu}

\begin{document}
\onecolumn
\firstpage{1}

\title[RR\,Lyrae stars with \textit{Kepler}]{RR\,Lyrae stars as seen by the \textit{Kepler} space telescope} 

\author[\firstAuthorLast ]{\Authors} 
\address{} 
\correspondance{} 

\extraAuth{}

\maketitle

\begin{abstract}

The unprecedented photometric precision along with the quasi-continuous sampling provided by the \textit{Kepler} space telescope revealed new and unpredicted phenomena that reformed and invigorated RR\,Lyrae star research.  The discovery of period doubling and the wealth of low-amplitude modes enlightened the complexity of the pulsation behavior and guided us towards nonlinear and nonradial studies. Searching and providing theoretical explanation for these newly found phenomena became a central question, as well as understanding their connection to the oldest enigma of RR\,Lyrae stars, the Blazhko effect. We attempt to summarize the highest impact RR\,Lyrae results based on or inspired by the data of the \textit{Kepler} space telescope both from the nominal and the K2 missions. Besides the three most intriguing topics, the period doubling, the low-amplitude modes, and the Blazhko effect, we also discuss the challenges of \textit{Kepler} photometry that played a crucial role in the results. The secrets of these amazing variables, uncovered by \textit{Kepler}, keep the theoretical, ground-based and space-based research inspired in the post-\textit{Kepler} era, since light variation of RR\,Lyrae stars is still not completely understood.



\tiny
 \keyFont{ \section{Keywords:} RR\,Lyrae stars, Kepler spacecraft, Blazkho effect, pulsating variable stars, horizontal-branch stars, pulsation, asteroseismology, nonradial oscillations} 
\end{abstract}

\section{Introduction}

RR\,Lyrae stars are large-amplitude, horizontal-branch pulsating stars which serve as tracers and distance indicators of old stellar populations in the Milky Way and neighboring galaxies.
They are also essential laboratories for testing evolutionary and pulsation models. Due to their importance and their large numbers among pulsating variables, they have been the subject of extensive research even before the era of space-based missions. Those studies had special interest in the properties of globular clusters, the pulsation period changes, and the long-standing mystery of the Blazhko effect \citep{blazhko-1907}. The Blazhko phenomenon is a quasi-periodic modulation of the pulsation amplitude and phase along with a prominent change of the light curve shape, which can be seen in a significant fraction of RR\,Lyrae stars. In spite of the enormous efforts both from theoretical and observational sides, the Blazhko effect is still not fully explained. However, except for this problem, RR\,Lyrae stars were thought to be a well-studied and well-understood class of pulsating variables before the launch of \textit{Kepler}. New phenomena hiding in the fine details were not expected. 

The dominant component of RR\,Lyrae light variation is the radial pulsation (with 0.2 to 1 day long periods) that can be either fundamental or first overtone mode, and in rarer cases these two modes appear simultaneously. The types are called RRab, RRc and RRd, respectively, following the tradition of Solon Bailey's original nomenclature \citep{bailey-1902}. The light curve shape itself is an excellent classifier of the pulsation mode and it is also useful to estimate physical parameters that make  RR\,Lyrae stars extremely valuable objects for asteroseismology (see the recent study of \citet{bellinger-2020} and references therein). By achieving the millimagnitude level in 
precision via space-based photometry in the last 20 years, it became clear that RR\,Lyrae pulsation is more complex than previously assumed, and in which nonradial pulsation and nonlinear dynamics also play important roles. 

Low-amplitude additional frequencies were first detected by the MOST space telescope \citep{matthews-2000} in the light curve of an RRd type star, AQ~Leo \citep{gruberbauer-2007}. 
No ground-based survey could compete with the continuity of space-based data that were eventually crucial in the discovery of the wealth of low-amplitude features in RR\,Lyrae stars. The CoRoT (Convection Rotation and planetary Transits) space mission \citep{baglin-2009} was launched almost three years before \textit{Kepler}, but by the time of the publication of the first RR\,Lyrae results, \textit{Kepler} had already been routinely delivering quarterly data. The first additional frequency in a CoRoT light curve was found by \citet{chadid-2010} in the Blazhko star V1127\,Aql, with a period ratio of $\sim$0.69 with the fundamental mode, which they also suggested to be a nonradial mode. This short prehistory of additional modes in RR\,Lyrae stars was then followed by the first \textit{Kepler} results and with these, a new era has begun.  

In this paper we present a summary of the most intriguing and defining discoveries in RR\,Lyrae stars achieved with or inspired by the \textit{Kepler} space telescope. We start with the 4-year long nominal mission and continue with the K2 mission which, in spite of technical difficulties, became surprisingly successful and equally important for RR\,Lyrae investigations.
At the time of this writing, \textit{Kepler} and especially, K2 data of RR\,Lyrae stars are far from being fully exploited, but two years after the official retirement of the telescope the time is ripe to review the major results obtained so far.

\section{The nominal \textit{Kepler} mission}

\textit{Kepler} was designed to find Earth-sized exoplanets around solar-like stars in the habitable zone with the transit method, hence at the core of its mission lies high-precision photometry of a large number of stars (on the order of $10^5$) for extended periods of time (originally planned for 3.5 years). Despite some minor difficulties (larger stellar noise than originally planned, unexpected safe-modes lasting for a few days to three weeks, the failure of one CCD module and later that of one of the reaction wheels), the mission was tremendously successful, providing an extraordinary wealth of planets and exotic planetary systems orbiting around a large variety of host stars.


The strategy in the original mission was to monitor one particular field of view (105 deg$^2$) close to the plane of the northern Milky Way that contains a large number of stars. In the telescope tube of the 95-cm effective-diameter Schmidt telescope a 42-CCD mosaic was collecting stellar light, slightly de-focused to shift the saturation limit and enable higher signal-to-noise per exposure. The individual exposures were approximately 6 seconds long, and were stacked to provide 1-minute (called short-cadence, for a small number of stars) and 30-minute integrations (referred to as long-cadence observations, the default observing mode for the vast majority of \textit{Kepler} targets). 
The cadence rate, along with the photometric precision and continuous nature of data, played an important role in \textit{Kepler}’s ability to improve on ground-based observations.

In March, 2009, \textit{Kepler} was launched to an Earth-trailing orbit around the Sun, with a period of 372.5 days. This ensured that the same part of the sky could be monitored continuously, as opposed to MOST, CoRoT, and the BRITE Constellation \citep{weiss-2014}, all moving on Earth-bound orbits.

In order to ensure optimal illumination of the solar panels of the spacecraft, a 90-degree rotation of the spacecraft was employed every 90-95 days (a quarter of the \textit{Kepler} year). The first two quarters (Q0, an engineering run lasting for 10 days, and Q1, an incomplete, 33-day long run) were performed keeping the same orientation, and after the first roll, Q2 was the first full-length quarter. The last quarter (Q17) meant a premature end of the original mission which was caused by the failure of a second reaction wheel. Reaction wheels were providing the stable and precise orientation of the spacecraft that played a crucial role in collecting close-to-micromagnitude precision light curves. 

Once every month during the mission (and eight times within 1.5 days at the beginning) full frame images (FFIs) were also taken. These were rarely used for scientific exploitation, although a few exceptions do exist \citep{montet-2016,hippke2018,molnar2018}. All \textit{Kepler} data, including raw RR\,Lyrae light curves and Full Frame Images are available at the MAST database\footnote{http://archive.stsci.edu/kepler/data\_search/search.php}.


Due to bandwidth limitations, only a small fraction of all pixels (typically 5-6\%) could be downloaded of \textit{Kepler}'s images. It meant that only pre-selected targets were observed, hence target selection prior to the main mission was crucial for the exoplanetary and also for the stellar astrophysics communities, although later on additional targets could be added through the Guest Observer opportunities. 
A thorough study to select main-sequence stars instead of giants was performed to help the achievement of the main mission goal, i.e., finding transiting exoplanets. The first and most important exoplanet results were announced by the core \textit{Kepler} Team \citep{borucki-2010}.

However, a smaller number (6000) of targets were dedicated to stellar characterization by the \textit{Kepler} Asteroseismic Science Consortium (KASC). Members of the KASC Working Group~\#13  which focused on RR\,Lyrae stars (later combined with WG\#7 for Cepheids, since the original \textit{Kepler} field contained only one classical Cepheid, V1154\,Cyg \citep{szabo-2011}), made sure that \textit{Kepler} observed all known RR\,Lyrae stars in the field. Therefore, all available variability catalogs were used for the pre-selection process. In total, roughly 50 RR\,Lyrae stars were found in the field prior to \textit{Kepler}'s launch. All targets that were found to be close to other targets were excluded, hence large contamination was expected. Specifically, the KASC RR Lyrae Working Group submitted 57 targets of which 21 were RRab stars (period range: 0.47-0.69 days, apparent \textit{Kepler} magnitudes ranging between 11.4–16.7), 2 RRc stars (V magnitude of 13.0 and 13.5 mag periods: 0.2485 and 0.3658 days), and all the rest were candidate RR Lyrae stars gathered from various ground-based variability surveys. It is worth noting that essentially no metallicity information was available for these targets. The situation was later remedied by \cite{nemec-2011} and \cite{nemec-2013}.

In addition, ground-based photometric measurements were also scheduled to find stars showing the Blazhko effect. 
Interestingly,  only one of targets among the RR Lyrae stars in the \textit{Kepler} field, namely RR Lyr itself had been known to be modulated, although at the time of launch we already knew that close to half of the RRab stars should show the phenomenon. The incidence rate was determined by a systematic survey of the Blazhko effect in a Galactic sample consisting of 30 RRab stars \citep{jurcsik-2009} and later was indeed confirmed by the \textit{Kepler} observations \citep{benko-2010}. 
This dramatic improvement in our knowledge of RR\,Lyrae stars just after a few weeks to months of \textit{Kepler} observations demonstrated the power of space photometric observations excellently. 

The quarterly roll of the spacecraft, however, caused discontinuities in the observations that led to the problem of data stitching. The position of the stars in the CCD changed with every roll, and the differences in pixel sensitivity appeared as mean brightness deviations as well as a bias in the pulsation amplitudes. This was complicated by the orbital motion of the satellite, the so-called \textit{Kepler} year that manifests itself in the differential velocity aberration, causing slow positional shifts of the targets on the CCDs. The extent of the shift is dependent on the positions in the field of view. To correct for these effects the type of the brightness variability had to also be taken into account. The larger the difference between the timescales of systematics and the intrinsic brightness variation, the easier to correct for them. 

The light curves could be detrended within the quarters by fitting and subtracting either a linear or polynomial fit, followed by a shift (addition) along with a stretch or a compression (multiplication) of the data. The zero point differences and the scaling factors with respect to the reference quarter (for that Q4 was often chosen) were straightforward  to calculate for the non-Blazhko stars, but not so simple for the Blazhko stars. Proper stitching of the modulated amplitudes needed special care, and the need to collect all the flux as accurately as possible, became essential. Therefore, tailor-made apertures were created for each target star and each quarter. These apertures contained all pixels that contained the signal of the star. This way the quarterly differences in the pulsation amplitudes were hoped to
practically disappear, leaving solely zero point shifts between the quarters. That could be not fully achieved, because the downloaded pixel `stamps' sometimes turned out to be too tight, creating inevitable flux loss. In these cases a few percent of scaling was applied to normalize the amplitudes from quarter to quarter \citep{benko-2014}.

The prototype star, RR Lyr was an important target of the mission. This star is not only the eponym of the type, but by far the brightest representative of its class, thus it saturated the \textit{Kepler} CCDs. With RR Lyr we learned an important lesson about how to collect all available flux and restore the full pulsation amplitude. High-amplitude variables can suffer significant flux loss at maximum light, if inadequate number of pixels are used in the photometry, and flux spilling along the pixel columns exceeds the aperture, which then distorts the measured light curve shape. The situation, when the central saturation column bleeds out from the aperture could be handled with a clever trick: the ratio of the central column flux to the adjacent column fluxes can be determined, and we can predict the flux when the central column is not fully captured. The light curve of RR Lyr was restored this way for quarters Q1--Q2 \citep{kolenberg-2011}.

\subsection{The discovery of period doubling and additional modes}

\begin{figure}[h!]
\begin{center}
\includegraphics[width=175mm]{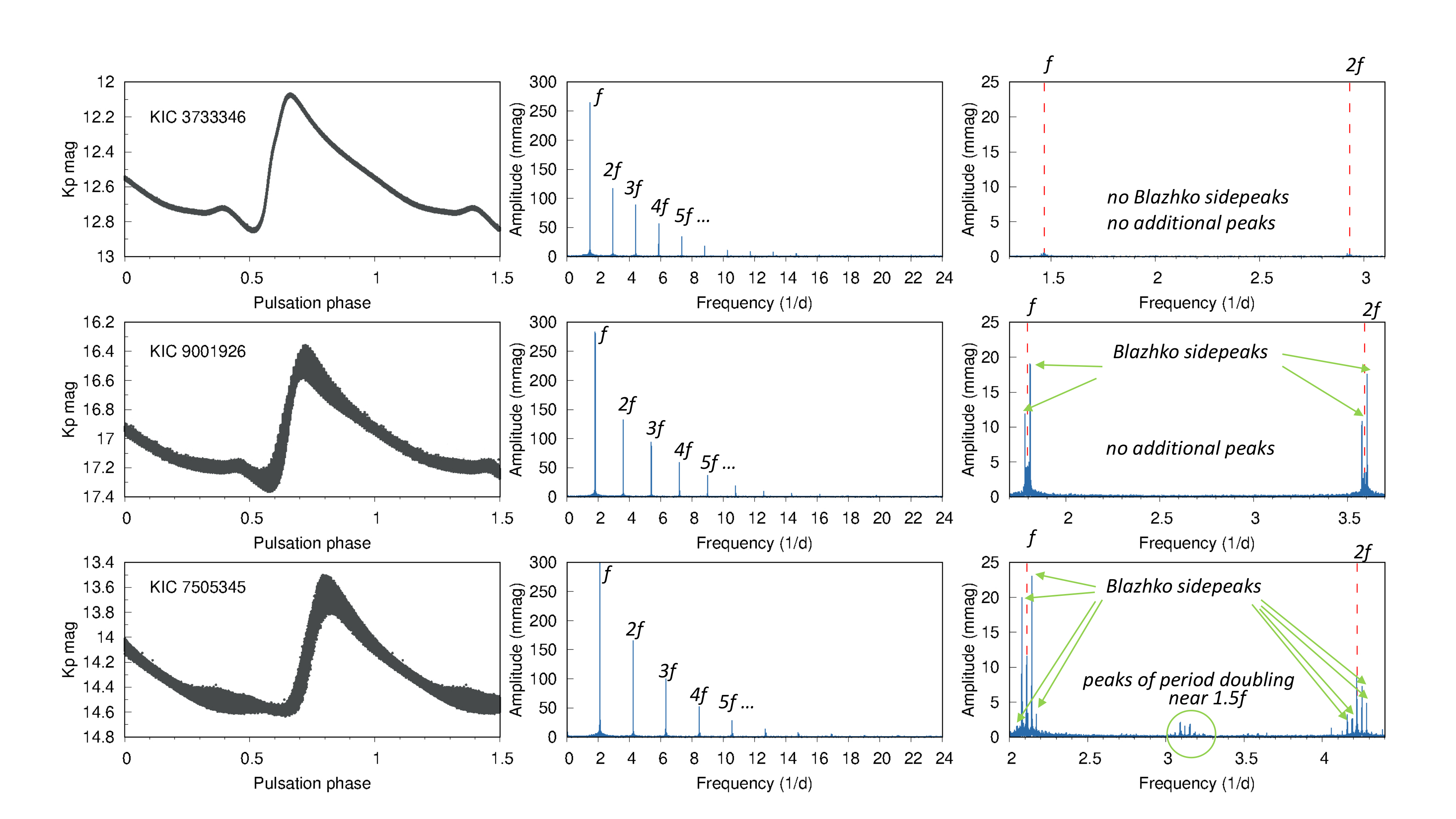}
\end{center}
\caption{Examples of non-Blazhko, Blazhko and period doubled Blazhko RRab stars of the \textit{Kepler} field. Panels show folded light curves, Fourier spectra, and zooms of residual spectra (after prewhitening with the pulsation frequency series: $f,2f,3f...$), respectively. Based on the light curves published by \citet{benko-2014}. }
\label{fig:new_magyarazo}
\end{figure}

The light curves of the first quarter (Q1) were analysed soon after they were obtained \citep{kolenberg-2010}. The unprecedentedly precise and continuous data were not only impressive, but clearly revealed that a much more complex pulsation is going on in RR\,Lyrae stars than we ever thought, and that the Fourier spectra displayed low-amplitude additional frequencies. The most surprising discovery among them was undoubtedly the half-integer (subharmonic) frequencies being present in some Blazhko RRab stars (at $0.5f$,$1.5f$, $2.5f$ etc., where $f$ is the fundamental-mode frequency). Fig. \ref{fig:new_magyarazo}. illustrates the differences between the spectra of a non-Blazhko RRab, a Blazhko RRab, and a Blazhko star with period doubling. It is important to mention here that only Blazhko RRab stars showed half-integer frequencies, RRc and non-Blazhko RRab variables did not, notwithstanding that the \textit{Kepler} sample size was limited. Namely, in Q1 42 RR Lyrae candidates were observed by \textit{Kepler} altogether: 17 Blazhko-modulated RRab, 17 non-modulated RRab, 4 RRc stars and 4 candidates turned out to be non RR Lyraes. 

\begin{figure}[h!]
\begin{center}
\includegraphics[width=160mm]{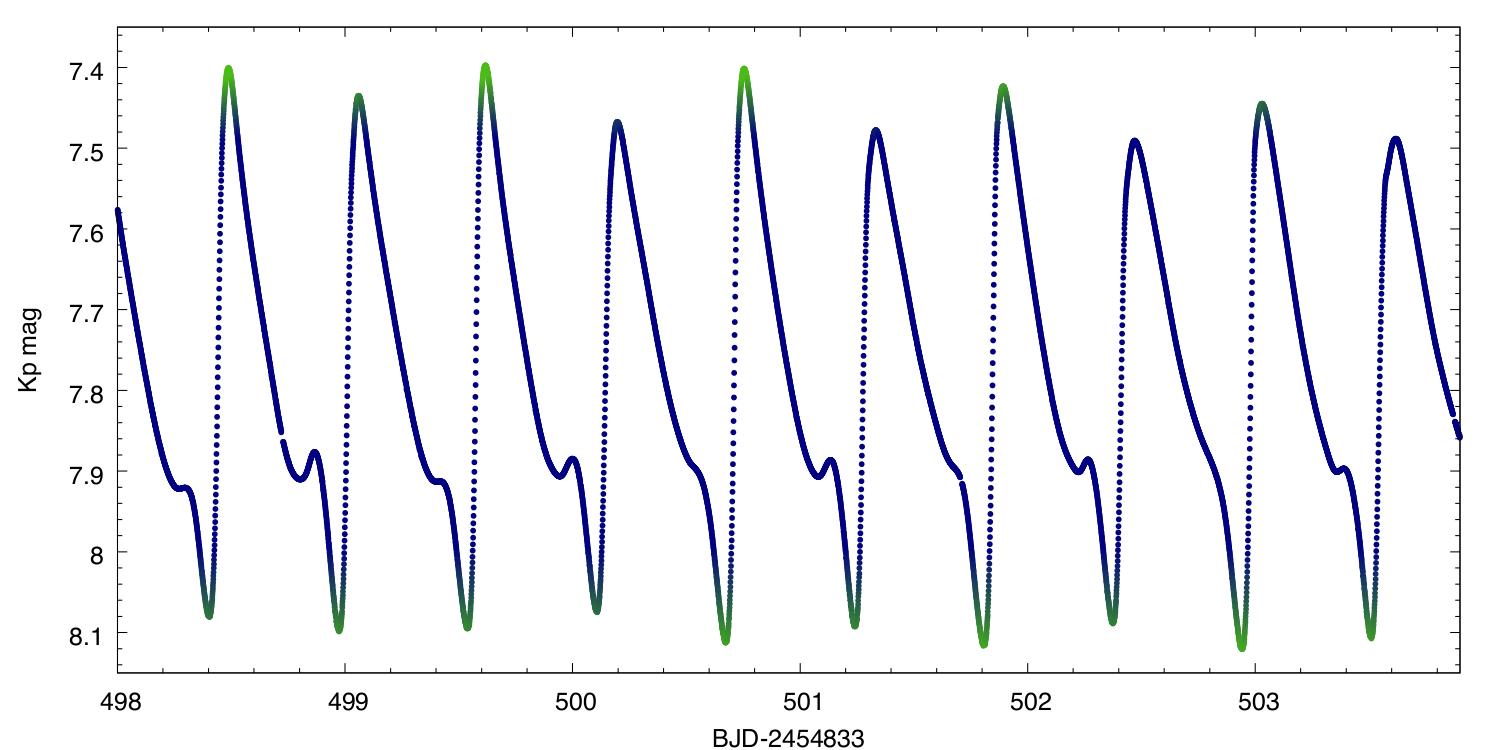}
\end{center}
\caption{Alternation of the pulsation amplitudes of RR~Lyr, the brightest representative of its class, as observed by \textit{Kepler}.  The observed fluxes were transformed to Kp magnitude, which is  \textit{Kepler}'s photometric system. Colors change with brightness for better visibility of the alternation. Data downloaded from the MAST.}
\label{fig:1}
\end{figure}

The half-integer frequencies were immediately recognized as the sign of period-doubling bifurcation, a nonlinear phenomenon, where the singly-periodic oscillation is destabilized by a resonance and turns into a two-period oscillation. The doubled period is visible in the light curves in the form of alternating low- and high-amplitude cycles of the original pulsation period (see Fig. \ref{fig:1}). This again showed the benefit of uninterrupted observations, since ground-based observations were inadequate to detect the period doubling with a characteristic timescale of close to one day, since on consecutive nights only every second pulsation cycles can be observed in a typical 0.5-day period RRab star.

\begin{figure}[h!]
\begin{center}
\includegraphics[width=175mm]{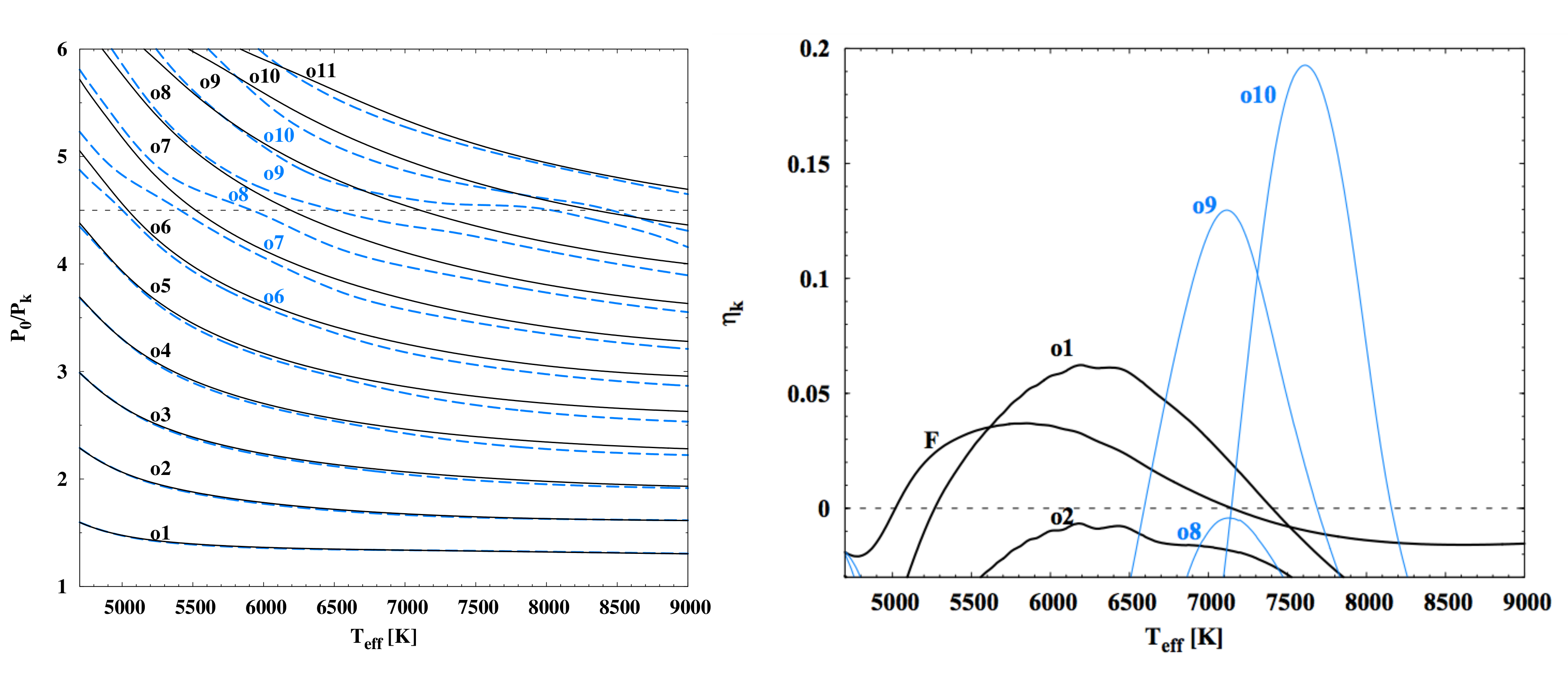}
\end{center}
\caption{Left: Modal diagram of an RR Lyrae model sequence \citep{kollath-2011}. Period ratios of adiabatic modes (black lines) and 
non-adiabatic modes (blue dashed lines). The 9/2 period ratio is indicated with the horizontal dashed line. Right: Linear growth rates of radial modes.  Overtones nine and ten could become excited.
Figure reproduced from Figures 3 and 4 of \citet{kollath-2011}.}\label{fig:new_model}
\end{figure}

By a fortunate coincidence, at about the same time period doubling was discovered in new RR\,Lyrae hydrodynamic simulations, providing an explanation and identifying the resonance that can be responsible for it \citep{szabo-2010, kollath-2011}. High-order radial overtones were investigated in detail and it was found that the ninth overtone can lock into a 9:2 resonance with the fundamental mode in a wide temperature range. Such high overtone modes are normally heavily damped and were thought to have no effect on the pulsation, but in RR\,Lyrae models this mode is trapped between the partial ionisation zone and the stellar surface (this type of mode is called a `strange mode'). The left panel of Fig. \ref{fig:new_model} shows the period ratios of the first eleven radial modes of an RR Lyrae model as a function of the effective temperature. Periods are scaled with the fundamental mode. Deviations in the nonadiabatic period ratios are most pronounced around 4.5, indicating the presence of strange modes. The right panel shows the linear growth rates of the same radial modes along with that of the fundamental mode. Instead of damping one finds excitation around the 9th and 10th radial overtones. To find out which high overtone is coupled to the fundamental mode a thorough nonlinear model search was performed in \cite{kollath-2011}, which resulted in the unambiguous detection of the 9:2 resonance between the fundamental mode and the 9th radial overtone. Note that the 9th overtone being locked in a 9:2 resonance is a pure numerical coincidence. The ability of inducing period doubling and maybe modulation (see later) by a high radial overtone is a strong effect that was completely unexpected prior to \textit{Kepler}.  


The unexpected discovery of period doubling was followed by the exploration of oscillations in the low-amplitude regime. 
\citet{benko-2010} analysed 29 RR\,Lyrae stars observed during the first 138 days of the mission (quarters Q0-Q2). Fourteen of them showed the Blazhko effect with modulation periods ranging from 28 days to much longer than the observing period. Beyond the usual ingredients of an RR\,Lyrae Fourier spectrum, namely the fundamental mode frequency and its harmonics plus the modulation multiplets around them, and the Blazhko frequency (occasionally along with its harmonics), new, low-amplitude periodicities were discovered at the millimag level and below. In four RRab stars these new frequencies fall close to the expected first or second (in some cases both) overtone pulsation frequencies. Three of these stars were identified as Blazhko-modulated ones, and one of them, V350\,Lyr was later recognized as the record-holder RRab with the smallest detected multi-periodic Blazhko-modulation ever (0.6 and 0.8 mmag, respectively, see \citealt{benko-2015}.)

Interestingly, linear hydrodynamic pulsation model computations presented in \citep{benko-2010} demonstrated that the fundamental mode and the second overtone can be simultaneously excited in these stars, so these observational results can be interpreted as the second radial overtone mode being excited with unusually low amplitudes, i.e. much lower than amplitudes of secondary frequencies in canonical double-mode RR Lyrae stars, where the amplitudes of the primary and secondary modes are comparable. Here there is a 2--3 orders of magnitude difference. A confirmation of this hypothesis can come from nonlinear pulsation simulations, but nonlinear effects both affect the mode selection process and shift the linear periods, and multi-mode pulsations are notoriously hard to reproduce. 

In fact, such a help from the modelling side seems to be inevitable, since theory suggests that periodicities close to radial overtone mode frequencies may arise from not only the radial modes themselves, but also from the dense spectrum of nonradial modes, that are preferentially excited close to the radial ones \citep{dziembowski-1977, van_hoolst-1998}. It is extremely interesting in this context that based on \textit{Kepler} observations, \citet{molnar-2012} found the first overtone to be excited with (very) small amplitude (at the 2-mmag level) in RR\,Lyrae, the prototype of the class. In fact – since RR\,Lyr is a Blazhko RRab, showing the period doubling phenomenon as well – one may assume that this star pulsates in three radial modes: fundamental mode, ninth radial overtone, and the first overtone. This was clearly demonstrated with the use of nonlinear hydrodynamical calculations, since the three-mode state showed up in the models as a stable state. Such model calculations, however, do not exist for the second radial overtone yet.

\subsection{The Blazhko effect seen in the \textit{Kepler} field }
The first analysis of the \textit{Kepler} data of RR Lyr was presented 
by \citet{kolenberg-2011}. The early data contained 
three Blazhko cycles only, but still clearly showed that the repetition was not strict. A slow shortening of the Blazhko period was documented before \textit{Kepler} \citep{kolenberg-2006}, but the new \textit{Kepler} data revealed that the Blazhko effect was variable on an even shorter time scale. At the same time, period doubling was also detected in RR Lyr. 

\begin{figure}[h!]
\begin{center}
\includegraphics[width=175mm]{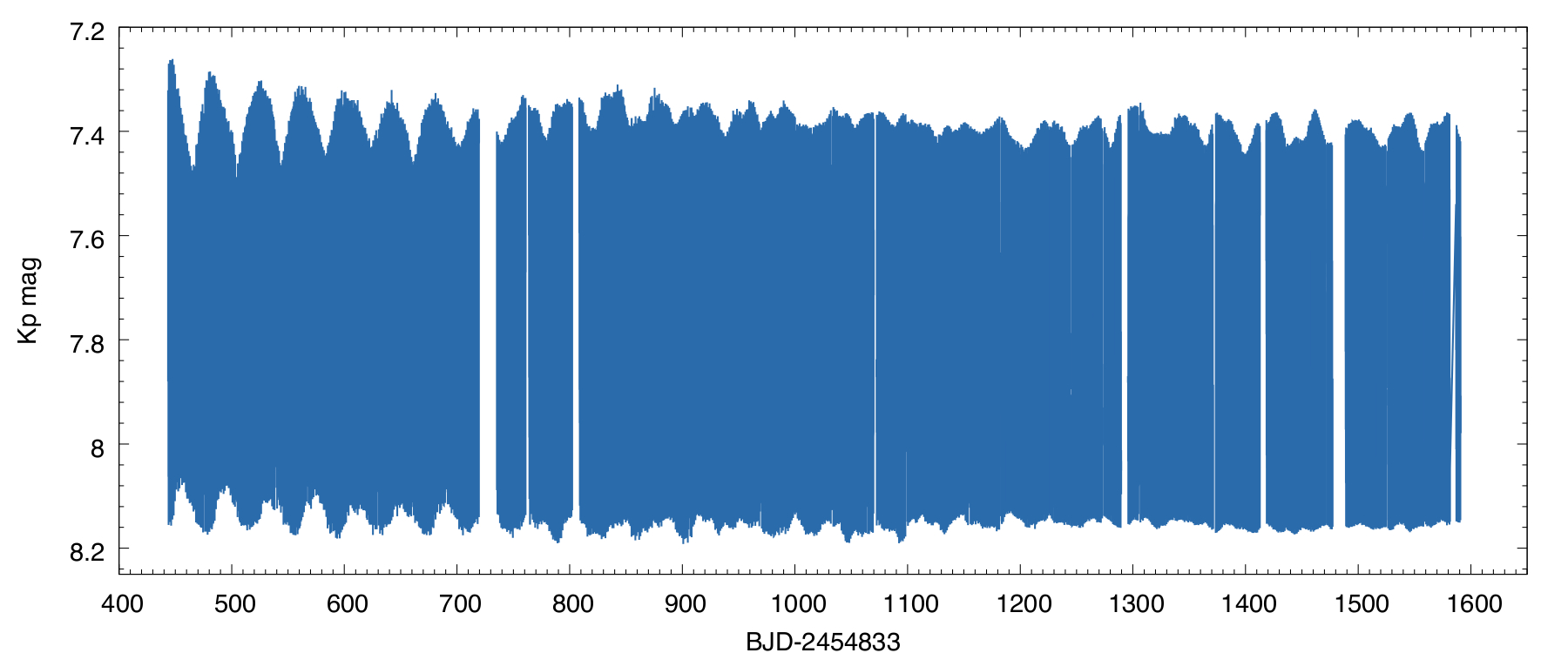}
\end{center}
\caption{The \textit{Kepler} Q5-Q17 short cadence data light curve of RR~Lyr, the eponym of its class. Data downloaded from the MAST. Note that individual pulsation cycles are not discernible on this scale. Gaps are present in the data due to safe mode events and other occasional technical problems.}\label{fig:2}
\end{figure}

RR Lyr was not observed in quarters Q3 and Q4, due to its underestimated brightness in the standard aperture assignment algorithms. Fortunately, the problem was  fixed, new custom apertures were set, and RR Lyr was observed until the end of the mission in short cadence mode (Fig. \ref{fig:2}). This rich data set witnessed a vanishing Blazhko effect \citep{stellingwerf-2013, leborgne-2014}. The Blazhko amplitude variation was 40\% of the pulsation amplitude in its strongest phase, but then decreased below 10\%, while the Blazhko period showed a 2\% decrease. This intriguing feature requires continuous monitoring, and RR Lyr provides a unique laboratory to study drastic changes in the pulsation state within human lifetime. 
 
The value of amplitude-independent methods is especially high for the analysis of the Blazhko modulation, due to the uncertainties in the pulsation amplitudes caused by the potential flux loss. The continuity and cadence of \textit{Kepler} data allow us to construct precise O--C diagrams that provide us not only with spectacular visualisations for the phase modulation, but they can be the subject of Fourier analysis themselves.

Fifteen Blazhko stars were found in the \textit{Kepler} field and analysed by \citet{benko-2014}. An unprecedentedly high percentage (80\%) of this sample was found to be multi-periodically modulated. Moreover, some of these stars showed different modulation periods to be dominant in the phase and in the amplitude variations. The ratio between the primary and secondary modulation periods was also investigated and found to be close to small integer numbers in almost all cases, suggesting that undiscovered resonances may play role in the modulation.

\begin{figure}[h!]
\begin{center}
\includegraphics[width=175mm]{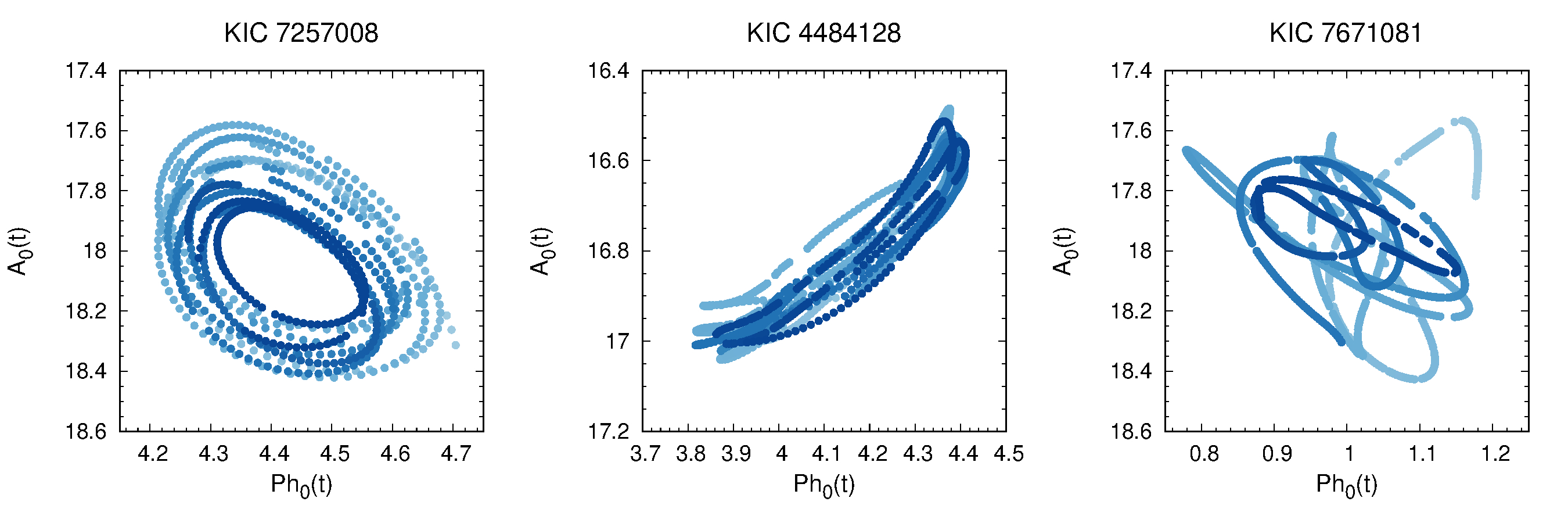}
\end{center}
\caption{Examples of the relation between the variations of the amplitude and phase of the pulsation frequency $f_0$ over time in three Blazhko stars. Color represents the progression of time, from light blue towards dark blue. Based on the light curves published by \citet{benko-2014}.}\label{fig:3}
\end{figure}

The Blazhko stars with monoperiodic modulation also showed some kind of irregularity. The nature of the irregularity (i.e., chaotic or stochastic) provides constraints for the theoretical models of the Blazhko effect, therefore V783 Cyg was investigated with nonlinear dynamical methods \citep{plachy-2014}. This star was the most promising candidate for that sensitive analysis, since it has the shortest modulation period, thus the number of observed modulation cycles during the \textit{Kepler} mission was the highest. The nonlinear dynamical analysis has strict requirements, only precise, continuous, and long input data (preferably of hundreds of cycles) will give reliable results. We were never able to collect such data for the Blazhko modulation in the pre-\textit{Kepler} era. The phase-space reconstruction applied to V783 Cyg revealed low-dimensional deterministic chaos in the dynamics behind the Blazhko modulation. However, the effect of instrumental issues was also tested, and it was found that the technical problems of the data stitching, the detrending and the sparse sampling may lead to apparent cycle-to-cycle variations in the modulation that could mimic the chaotic behaviour. Thus the intrinsic origin of the irregularity in the Blazhko effect could not be proven for V783 Cyg, showing the limitations of \textit{Kepler} data.  

Studying the relation between the phase modulation and the amplitude modulation we can discover various morphology types. In Fig. \ref{fig:3} we present three examples (KIC~7257008, KIC~4484128, KIC~7671081) where instantaneous amplitudes of the $f_0(t)$ pulsation frequency are plotted as the function of the instantaneous phase. These trajectories show very different routes, for which no explanation or connection to other pulsation properties has been found yet.

The connection between the Blazhko effect and the period doubling phenomenon is also an important question that need to be investigated since only the Blazhko stars show the period doubling. Nine out of the 15 Blazhko stars of the \textit{Kepler} field show half-integer frequencies, while the others do not. The half-integer frequency series appears with the highest peak typically at $1.5f$, which is very often a non-coherent peak suggesting temporal variability. Indeed, the alternation of the pulsation cycles in the light curves sometimes disappears or decreases to a very low amplitude. The order of the low- and high-amplitude cycles also changes, this feature was visualised by \citet{molnar-2014} who connected every second maxima for the even and for the odd cycles with separate lines. The interchanges in the order occur when these lines cross each other. Similar interchanges were observed in the pulsation of RV\,Tau stars \citet{plachy-interchange}. 

\subsection{The wealth of low-amplitude frequencies}

\begin{figure}[h!]
\begin{center}
\includegraphics[width=170mm]{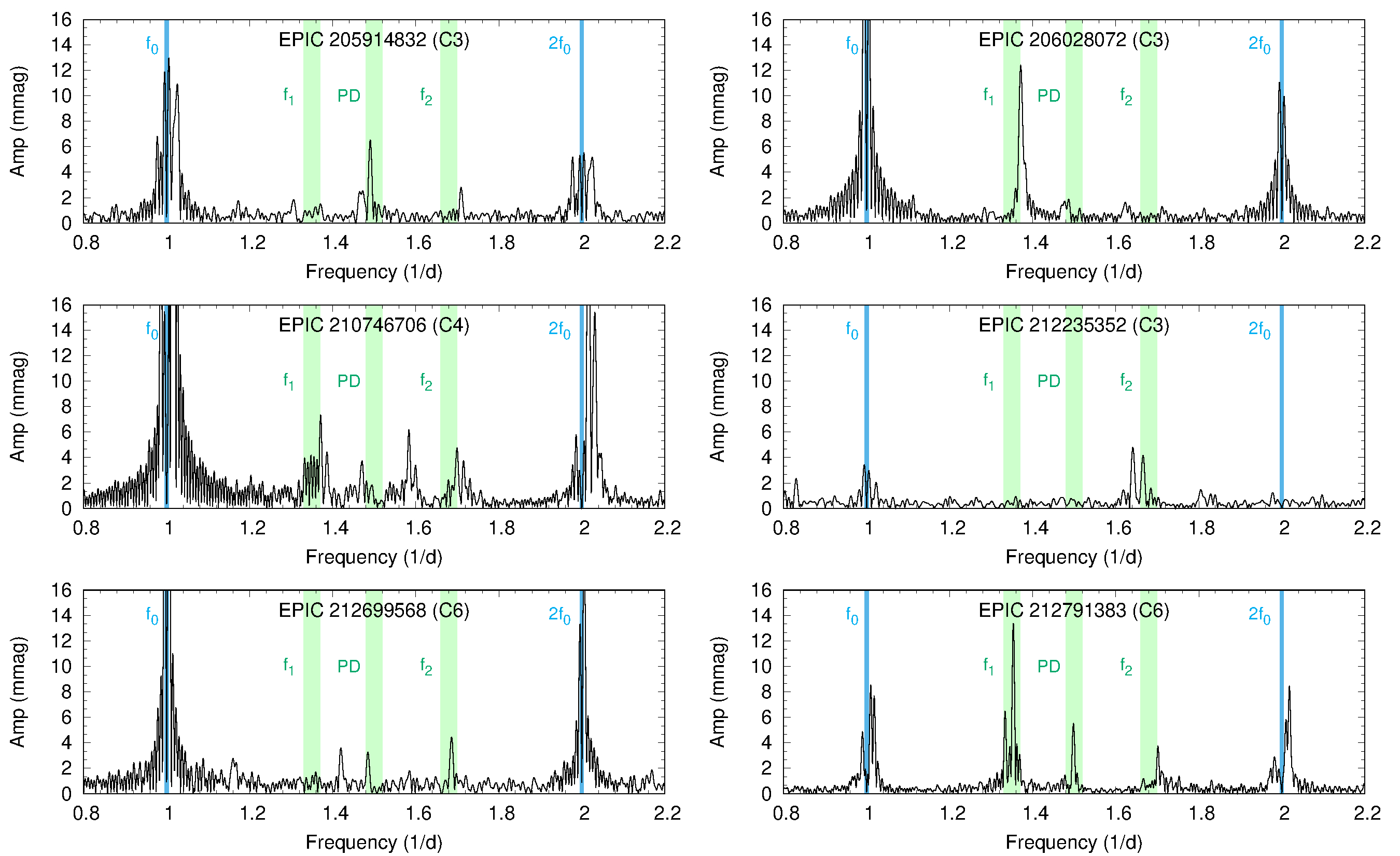}
\end{center}
\caption{Examples of residual Fourier spectra of K2 RR\,Lyrae stars showing low-amplitude additional peaks (after removing the frequency series of the main pulsation). Green areas are the expected frequency regions of the first radial overtone, period doubling (PD) and the second radial overtone,  respectively. Based on the light curves published by \citet{eap}.}\label{fig:4}
\end{figure}

The detailed Fourier analysis of the Blazhko RRab stars recovered different groups of low-amplitude additional frequencies \citep{benko-2014}. The expected regions of the first ($f_1$) and the second overtone ($f_2$) are positioned at the two sides of the $1.5f$ frequency group in the Fourier spectra. Frequencies at millimagnitude level appear within these regions, but sometimes also between them. We demonstrate this on examples from the K2 mission in Fig. \ref{fig:4}. The origin of frequency peaks outside the expected frequency regions is unclear.

A rich frequency spectrum of additional modes was observed in V445 Lyr, a Blazhko star showing extreme strong modulation \citep{guggenberger-2012}. The peaks belonging to the period doubling, the first and second overtones all appeared, the latter with a clearly variable amplitude that was not connected to the Blazhko phase. A fourth peak was interpreted as a nonradial mode. Altogether 80 combination frequencies have been identified in this star.

The analysis of the four \textit{Kepler} RRc stars resulted in the clear detection of low-amplitude oscillations with period ratio of $P/P_1$ = 0.612--0.632 with the dominant first overtone mode \citep{moskalik-2015}. Subharmonics of these $f_{0.61}$ frequencies at 0.5$f_{0.61}$ and 1.5$f_{0.61}$ have also been detected. This type of low-amplitude modes were identified in Cepheid stars in the Large Magellanic Cloud, all within similar period ratios (0.6--0.64) relative to the overtone mode \citep{moskalik-2008}. Nonradial oscillation was immediately proposed as the origin. This discovery was made possible by the OGLE (Optical Gravitational Lensing Experiment) survey that became a fundamental source of high-quality ground-based photometry of radial pulsators in the last decade. 
Modes with similar period ratios were first seen in the RRc type by \citet{olech-2009} among the RR\,Lyraes of the $\omega$~Centauri. Since then, it became clear that these stars form similar sequences in the Petersen diagram as overtone Cepheids do \citep{smolec-2017}. A model explaining the nature of these additional periodicities has been proposed by \citet{dziem-2016}, in which 
strongly trapped, unstable nonradial modes ($\ell$ = 7, 8 and 9 degrees in classical Cepheids and $\ell$ = 8 and 9 in RR Lyr stars) are excited. In this model the nonradial mode is actually at the $0.5f_{0.61}$ frequency, while its first harmonic signal can reach higher amplitudes due to geometric and nonlinear effects.

\begin{table}
\centering
\caption{Additional modes in the \textit{Kepler} RR Lyrae stars.} \label{tab:targets}
\begin{tabular}{rllll}

KIC number & GCVS name & Subtype & Period & Additional frequencies  \\

3864443	&	V2178 Cyg	&	RRab-BL	&	0.486947	&	$f_2$, PD	\citep{benko-2010}					\\
4484128	&	V808 Cyg	&	RRab-BL	&	0.5478635	&	PD	\citep{benko-2010},			\\
	&	&	&		&	$f_2$	\citep{benko-2014}		\\
5559631	&	V783 Cyg	&	RRab-BL	&	0.6207001	&							\\
6183128	&	V354 Lyr	&	RRab-BL	&	0.5616892	&	$f_1$, $f_2$, PD	\citep{benko-2010},				\\
&		&	&		&	(no $f_1$)	\citep{benko-2014}		\\
6186029	&	V445 Lyr	&	RRab-BL	&	0.5130907	&	$f_1$, $f_2$, PD	\citep{benko-2010}					\\
7198959	&	RR  Lyr	&	RRab-BL	&	0.566788	&	PD	\citep{benko-2010}, 			\\
&	&		&		&	 $f_1$	\citep{molnar-2012}					\\
7505345	&	V355  Lyr	&	RRab-BL	&	0.4736995	&	PD	\citep{benko-2010},		\\
&		&		&		&			$f_2$	\citep{benko-2014}		\\
7671081	&	V450  Lyr	&	RRab-BL	&	0.5046198	&	$f_2$	\citep{benko-2014}					\\
9001926	&	V353  Lyr	&	RRab-BL	&	0.5567997	&							\\
9578833	&	V366  Lyr	&	RRab-BL	&	0.5270284	&	$f_2$	\citep{benko-2014}					\\
9697825	&	V360  Lyr	&	RRab-BL	&	0.5575755	&	$f_1$, PD	\citep{benko-2010},			\\
	&		&	&		&	$f_2$ (no $f_1$)	\citep{benko-2014}		\\
11125706	&		&	RRab-BL	&	0.61322	&							\\
12155928	&	V1104 Cyg	&	RRab-BL	&	0.4363851	&							\\
7257008	&		&	RRab-BL	&	0.511787	&	$f_2$, PD 	\citep{benko-2014}					\\
9973633	&		&	RRab-BL	&	0.510783	&	$f_2$, PD 	\citep{benko-2014}					\\
10789273	&	V838 Cyg	&	RRab-BL	&	0.48028	&	$f_2$, PD 	\citep{benko-2014}			\\
9508655		&	V350 Lyr	&	RRab-BL & 0.59424&$f_2$	\citep{benko-2010}	\\
7021124		&		&	RRab-BL & 0.6224925& $f_2$ \citep{nemec-2011}\\	\\
3733346	&	NR Lyr	&	RRab	&	0.6820264	&							\\
3866709	&	V715 Cyg	&	RRab	&	0.47070609	&							\\
5299596	&	V782 Cyg	&	RRab	&	0.5236377	&							\\
6070714	&	V784 Cyg	&	RRab	&	0.5340941	&							\\
6100702	&		&	RRab	&	0.4881457	&							\\
6763132	&	NQ Lyr	&	RRab	&	0.5877887	&	$f_1$	\citep{benko-2019}					\\
6936115	&	FN Lyr	&	RRab	&	0.52739847	&							\\
7030715	&		&	RRab	&	0.68361247	&							\\
7176080	&	V349 Lyr	&	RRab	&	0.507074	&							\\
7742534	&	V368 Lyr	&	RRab	&	0.4564851	&							\\
7988343	&	V1510 Cyg	&	RRab	&	0.5811436	&	$f_2-f_0$	\citep{benko-2019}					\\
8344381	&	V346 Lyr	&	RRab	&	0.5768288	&	$f_2-f_0$	\citep{benko-2019}					\\
9591503	&	V894 Lyr	&	RRab	&	0.5713866	&	$f_2-f_0$	\citep{benko-2019}					\\
9658012	&		&	RRab	&	0.533206	&	$f_2$	\citep{benko-2019}					\\
9717032	&		&	RRab	&	0.5569092	&							\\
9947026	&	V2470 Cyg	&	RRab	&	0.5485905	&	$f_1$	\citep{benko-2019}					\\
10136240	&	V1107 Cyg	&	RRab	&	0.5657781	&							\\
10136603	&	V839 Cyg	&	RRab	&	0.4337747	&							\\
11802860	&	AW Dra	&	RRab	&	0.687216	&							\\
														\\
8832417	&		&	RRc	&	0.2485464	&	$f_{0.61}$	\citep{moskalik-2015}					\\
5520878	&		&	RRc	&	0.2691699	& $f_{0.61}$	\citep{moskalik-2015}					\\
4064484	&		&	RRc	&	0.3370019	&	$f_{0.61}$	\citep{moskalik-2015}					\\
9453114	&		&	RRc	&	0.3660809	&$f_{0.61}$, $f_{0.68}$	\citep{moskalik-2015}			\\		
														
\end{tabular}
\label{table:targets}
\end{table}

Nineteen non-modulated RR Lyrae stars were analysed by \citet{nemec-2011}. None of these stars show the period doubling phenomenon that strongly suggests its connection to the Blazhko effect. Empirical photometric metal abundances were also derived for these stars and compared to spectroscopic metallicities. The so-derived metallicities of most of the stars were found to be similar to those of intermediate-metallicity globular clusters ([Fe/H]$\sim$ --1.6). However, the lowest-amplitude stars turned out to be metal-rich with [Fe/H] between --0.55 and +0.07. Spectroscopic observations of the \textit{Kepler} field RR\,Lyrae stars were performed by the Canada--France--Hawaii 3.6-m telescope (CFHT) and the Keck-I 10-m telescope (W.M. Keck Observatory) \citep{nemec-2013}, and these measurements confirmed  the photometric [Fe/H] results for the non-Blazhko and most Blazhko RRab stars.

The reanalyis of the non-Blazhko stars revealed cycle-to-cycle light curve variations in stars which are brighter than Kp$\sim$15.4 mag \citep{benko-2019}. Scattered short-cadence  observations that have been obtained for a few quarters for non-Blazhko \textit{Kepler} RR\,Lyrae stars were crucial in this discovery. The amplitude differences between the light curve maxima were found to be in the range of 5--8 mmag. Additional modes identified as the first and the second overtone modes have also been recovered at several non-Blazhko stars with extremely low amplitude ratios with the fundamental mode. Moreover, the amplitude of the additional modes show changes over time, while in some cases they appear only temporarily. Their linear combinations with the fundamental mode were also detectable, sometimes with much higher amplitude than the low-amplitude additional modes themselves. This again suggests the nonradial origin of these modes, since a scenario was found only in case of nonradial modes. In Table~\ref{table:targets}. we summarized the identification of low-amplitude additional frequencies in the \textit{Kepler} RR Lyrae stars for which detailed Fourier analysis has already been performed. 

The \textit{Kepler} data of non-Blazhko stars provide opportunity to search for binarity too. The variation of the O--C diagrams of two RRab stars could be fitted with the light time effect caused by a low-mass companion, likely a substellar object (giant planet or brown dwarf) \citep{li-2014}. Comparing the high number of RR\,Lyrae stars and the few binary candidate systems that have been found so far \citep{hajdu-2015, skarka-2019}, we may conclude that RR\,Lyrae close-in companions are very rare, in accord with (binary) stellar evolution theory predictions of stars that are past the red giant phase. 

\subsection{Studies inspired by \textit{Kepler} RR Lyrae results}

The \textit{Kepler} RR\,Lyrae results entirely reformed the research field of classical pulsators. Searching for more examples of period doubling and nonradial modes in hydrodynamical models and high-quality ground- and space-based photometry became the most relevant.   

The discovery of period doubling revitalized the theoretical side of the radial stellar pulsation field.
Hydrodynamic models of the '90s already predicted that period doubling can naturally occur in Cepheids \citep{moskalik-buchler-1990}. Cepheids are radially pulsating siblings of RR\,Lyrae stars showing pulsation behavior similar in several aspects, most prominently in the shape of the light curve. These stars, however, are more massive and cross the classical instability strip at different evolutionary phases. The two main types are distinguished based on their Population I or II membership, and this also determines the nonlinear phenomena they can exhibit. The weakly dissipative Population I Cepheid models showed only period doubling, whereas models of the strongly dissipative Population II Cepheids followed a cascade of period doubling bifurcation that eventually led to chaos. The destabilisation was caused by low-order half-integer resonances at both types  (3:2 and 5:2, respectively).  

As the period doubling appeared in the RR\,Lyrae models \citep{kollath-2011}, the suspicion that resonant interaction between the fundamental mode and ninth overtone might also be the key in the Blazhko immediately arose. The hydrodynamical simulations, however, did not produce modulation. Only the amplitude equation formalism provided a demonstration that amplitude modulations may occur as a result of nonlinear, resonant mode coupling between these two modes \citep{buchler-2011}. This new theory of the Blazhko effect predicts both type of modulations: periodic and chaotic. 

The resonant mode coupling mechanism is also supported by BL\,Her hydrodynamical calculations \citep{smolec-2012a}. The models of these short-period Type II Cepheids exhibited periodic and quasi-periodic modulation of the pulsation amplitudes and phases. Moreover, some models showed period doubling with or without modulation. A 3:2 resonance was identified behind the phenomenon. This theoretical work was then further improved and two period-doubling domains were recovered, one between 2-6.5 days, and one above 9.5 days, in the regime of W\,Vir-type stars \citep{smolec-2016}.

The luminous siblings of RR\,Lyrae stars, the classical Cepheids are the primary standard candles in extragalactic distance measurements, and they were believed to be clockwork-precision pulsators before the era of ultra-precise measurements. The only exception was the unique case of V473 Lyrae, the only known Cepheid displaying strong Blazhko effect \citep{burki-1980}. The pulsation of this star is multiply modulated and the measurement by the MOST space telescope also revealed period doubling in it \citep{molnar-2017}.

Regarding Type II Cepheids, cycle-to-cycle variations were known to be common among W\,Vir stars and towards the longer periods (P$>$20 d).  The RV\,Tau phenomenon was recognized as period doubling, caused most likely by a 2:1 resonance \citep{fokin-1994}. Nevertheless, the first detection of period doubling in W\,Vir stars has only been achieved with the \textit{Kepler} space telescope in the K2 mission \citep{plachy-2017}. In this new context it was also recognized that the amplitude alternation seen by \citet{templeton-2007} in multicolor observations of W\,Vir, the eponym of the class, is actually period doubling. Thanks to the extensive analysis of Cepheids in the Galactic bulge by the OGLE survey, we know that period doubling is common for periods above 15 days and the transition towards the RV\,Tau regime is a smooth process \citep{smolec-2018}. Period doubling at shorter periods has also been discovered in OGLE BL\,Her stars \citep{smolec-2012b}, but only three such stars are known so far, suggesting that period doubling is a rare phenomenon in that class. 

Based on the \textit{Kepler} findings the CoRoT RR\,Lyrae data were revisited to look for phenomena that might be overlooked before \citep{szabo-2014}. The most important result was the discovery of period doubling in modulated CoRoT RRab stars. Short sections of alternating maxima, typical of the period doubling effect, were found in four CoRoT RR\,Lyrae stars out of six modulated RRab stars. Given the usually brief time intervals where the phenomenon is detectable, occurrence can be even higher. This result corroborates with the previous claims about the strong correlation of the occurrence of period doubling and the Blazhko phenomenon, since no period doubling was found in non-modulated RRab or RRc/RRd stars in the CoRoT sample. It also became clear that additional modes are ubiquitous in all RR\,Lyrae subtypes except for the non-modulated RRab pulsators. In these stars no additional periodicities were found down to the precision of CoRoT and \textit{Kepler}. In addition, non-coherence (temporal variability) of the additional modes in RRc, RRd and modulated RRab stars were found to be dominant. While in the latter group the Blazhko modulation itself might be a clue, some other mechanism should be at work in the overtone and classical double-mode pulsators. 

\section{The K2 mission}

\textit{Kepler} was only barely able to extend its original mission. By May of 2013 only two functioning reaction wheels remained on the spacecraft out of four, thus tri-axial stabilization was lost. It became clear that the original mission could not be continued without significant deterioration in data quality. NASA called for ideas and methods for a new observational strategy in the two-wheel mode. The scientific community reacted fast to save the mission, 42 white papers have been submitted including two that considered potential RR\,Lyrae observations. One suggested a continued observation of the \textit{Kepler} field of view with extended number of high cadence targets among the large amplitude pulsators \citep{whitepaper-1}. Beside many advantages, the longer time span would have been allowed to study the Blazhko effect in more detail. The other white paper proposed to turn the telescope to the South Ecliptic Pole to monitor the largest possible sample of well classified large-amplitude pulsating and eclipsing variables \citep{whitepaper-2}. That would have meant synergy with the OGLE survey and the TESS mission.

The final concept opted for a solution that maximizes the photometric performance by minimizing the roll of the spacecraft. This was possible with target fields placed along the Ecliptic plane. Each field could be monitored for about 80 days, and before changing fields the telescope had to turn sideways to keep the fixed solar panels aimed at the Sun. These periods were called `Campaigns'. The design of the mission opened new opportunities for RR\,Lyrae investigations, as well as many other fields of astronomy. 
With the changing fields of view, massive continuous space photometry became accessible for a variety of stellar objects the first time. The new mission was named K2 \citep{howell-2014}.

The observation started with an 8.9-d long Two-Wheel Engineering Test in February of 2014, which targeted nearly two thousand stars. The K2-E2 sample contained 33 RR\,Lyrae stars, members from each subtypes (RRab, RRc, RRd). This short run demonstrated that all new low-amplitude phenomena seen in the \textit{Kepler} mission can be recovered from K2 data too \citep{molnar-2015b}. Period doubling was detected in two RRab stars, and nonradial modes in two RRd and the three non-modulated RRc stars. The first space-based photometry of a Blazhko-RRc star was also provided by K2 Engineering Test. 

With K2 we lost the chance to study stars showing long-period Blazhko effect, but in exchange we got a large sample of RR\,Lyrae stars from different regions of the Galaxy, providing us with a basis for population studies and statistical investigations. Careful target selection therefore became a new important task for the KASC Working Group~\#7.
The \textit{Kepler} Guest Observer Office developed the \texttt{K2FoV} tool to check the visibility of targets in each Campaign. Proposals could be submitted through the NASA NSPIRES System starting from Campaign 3. The existing large ground-based surveys, such as the Catalina Sky Survey \citep{catalina}, the  Lincoln Near Earth Asteroid Research
\citep{linear}, the All Sky Automated Survey \citep{asas}, and the Northern Sky Variability Survey \citep{nsvs} offered photometric data of a great number of RR\,Lyrae candidates that we subsequently used in the target selection \citep{plachy-2016}.
RR\,Lyrae proposals prioritized the less common double-mode and overtone RR\,Lyrae stars, and those RRab stars that looked special in some sense, like having extreme Blazhko modulation or an unusually long pulsation period. A few of the most interesting targets were also proposed for 1-minute cadence. The observable targets in the K2 mission were limited by the telemetry, 10 to 20 thousand long cadence targets and 50 to 100 short cadence targets were available per Campaign. Nevertheless, most of the proposed RR\,Lyrae targets have been approved during the K2 mission, (except for the first Campaigns), altogether reaching about 4000 RR\,Lyrae stars. At the time of writing this paper, only a small fraction of this huge sample has been analysed, leaving the better part of it for future studies. 
The major problem that slows down the massive analysis is the pointing jitter of the space telescope. 
No comprehensive correction solution could be invented so far for this problem, which would work well also for RR~Lyrae stars.
The EVEREST pipeline comes close but even that one removes the pulsation signal from a significant fraction of RR Lyrae stars \citep{luger2018}. The instrumental signal is similar to RR\,Lyrae light curves in periodicity and shape, which makes its elimination challenging. We report the discoveries based on early K2 RR\,Lyrae data in the following sections.

\subsection{The challenges of K2 RR Lyrae photometry}

The two-wheel mode of the K2 mission made attitude control maneuvers of the satellite necessary in about every 6 hours (occasionally 12 hours). These maneuvers corrected the the torque caused by the radiation pressure that distributed unevenly on the spacecraft. As a consequence, sudden drifts and jumps occurred in the positions of the field of view that reached as much as two pixels at the edges of the detector. Therefore, the sensitivity variation within and between the pixels caused a systematic variation in the light curves. In worse cases they were contaminated by the flux from nearby stars as well. Besides the Simple Aperture Photometry (SAP) and Presearch Data Conditioned SAP (PDCSAP) data products provided by the mission \citep{vancleve-2016}, several other pipelines have been developed to fix or at least minimize the instrumental effects. These pipelines used different approaches to find general solution to all variables, but most of them failed on RR\,Lyrae stars for two main reasons. Apertures were too tight to capture the $\sim$1 magnitude variability and/or the correction methods could not distinguish between the sharp features of systematics and RR\,Lyrae variation that occurs on the same timescale. A comparison analysis of various photometry outputs has been carried out by \citet{eap}, who suggested a method optimized for the detection of RR\,Lyrae light variation. The main idea for the EAP (Extended Aperture Photometry) method was the extension of the apertures to contain the star in the maximum brightness phases. This alone improved the light curve significantly and applying the K2SC (K2 Systematics Correction) pipeline \citep{aigrain-2016} on EAP solutions, light curves improved even more. Over four hundred such RR\,Lyrae light curves have been prepared and analyzed so far from Campaign 3 to 6. 

Anomalous Cepheids constitute a rare type of pulsating stars. They are 2-3 times more massive than RR\,Lyrae stars, and pulsate with periods ranging from 0.3-2 days either in the  fundamental or in the first  overtone. Their origin is not fully clear, probably both single-star and binary evolution channels contribute to the production of these objects \citep{bono,acep-evol}. RR\,Lyrae stars and anomalous Cepheids can be distinguished based on their light curve shapes, but only precise photometry can show their slightly distinct location in the Fourier parameter plane of $log(p)-\phi_{21}$ and $log(p)-\phi_{31}$. A useful byproduct of K2 RR\,Lyrae analysis was the discovery of four new anomalous Cepheid candidates \citep{eap}. All four candidates found among K2 RR\,Lyrae stars are fundamental mode pulsators, and they also provide the first detection of Blazhko-modulation in this variable type.   

\begin{figure}[h!]
\begin{center}
\includegraphics[width=170mm]{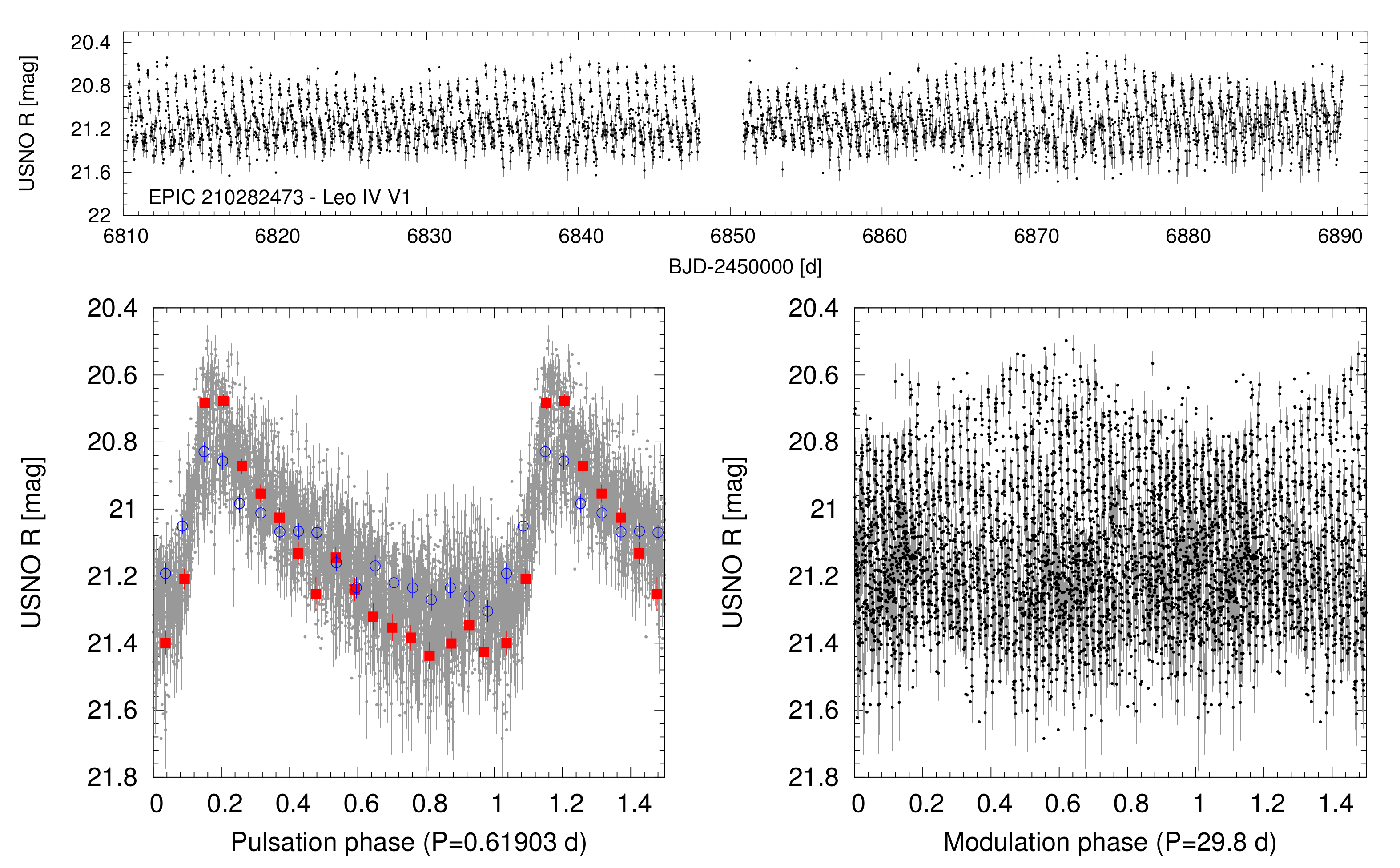}
\end{center}
\caption{The faintest Blazhko star from the K2 mission, EPIC 210282473 (Leo IV V1). Upper panel: the K2 light curve. Lower left panel: phase curve folded by the pulsation period, 2 day binned data from the maximum-amplitude phase (red) and minimum-amplitude phase (blue) are marked. Lower right panel: phase curve folded with the modulation period. Figure reproduced from Figures 3 and 5 in \citet{molnar-2015} \copyright{} AAS. Reproduced with permission.} \label{fig:new}
\end{figure} 
Campaign 1 pointed to the North Galactic Cap containing the dwarf spheroidal galaxy Leo IV in the field of view. Leo IV is one among the ultra-faint satellites of the Milky Way discovered by the Sloan Digital Sky Survey \citep{belokurov-2007}. Three fundamental-mode RR\,Lyrae have been identified in Leo IV by \citet{moretti-2009} at brightness $\sim$21.5 mag in $V$ band. All three RRab were detectable in K2 images, which made them the faintest pulsating stars measured with \textit{Kepler} ever \citep{molnar-2015}. To obtain accurate photometry for such faint objects, image subtraction techniques had to be involved. That was performed with the open source \texttt{FITSH} software package \citep{fitsh}. The individual frames had to be adjusted to the same reference system to compensate for the pointing motions. Additional neighbouring K2 stamps were used beside the target frames to determine the precise transformations and background level. This technique ensured light curve precision in the order of 50-100 mmag, and that was adequate to detect the farthest Blazhko modulation from Earth, as well as the first clear detection of Blazhko effect beyond the Milky Way and the Magellanic Clouds (Fig. \ref{fig:new}). The existence of Blazhko effect in such a metal poor galaxy as Leo~IV ($\langle [F/H] \rangle=\sim-2.3$) also provides constraints for the theoretic models, while the experience collected with this study will be useful for future investigations of faint and/or extragalactic objects of the K2 mission. 

\subsection{Characterization of the low-amplitude modes}

\begin{figure}[h!]
\begin{center}
\includegraphics[width=175mm]{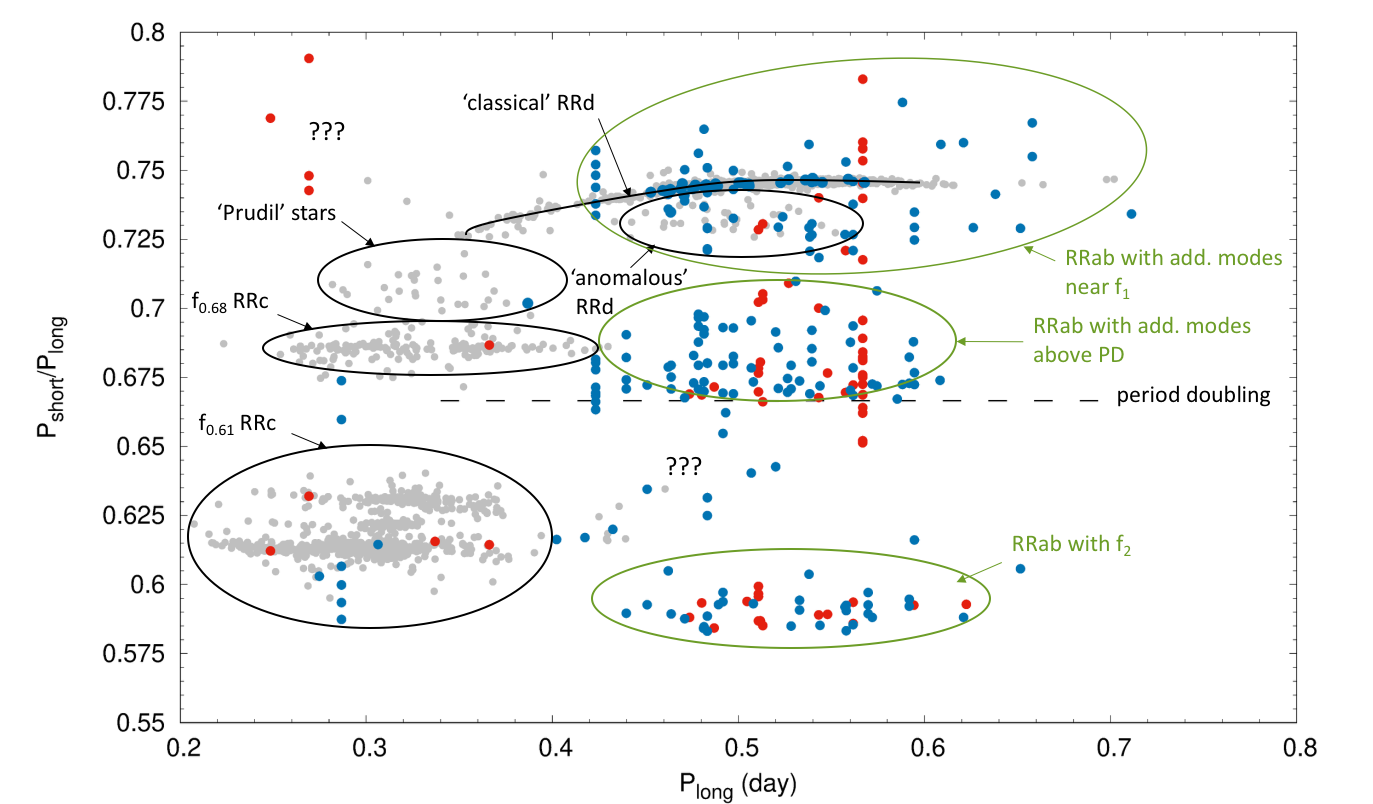}
\end{center}
\caption{Petersen diagram of multi-mode RR Lyrae stars. The grey points denote to OGLE discoveries collected from \citep{sos-2016, mc-rrl, sos-2019, prudil-stars, netzel-2019}. Red points are results from the \textit{Kepler} field \citep{moskalik-2015, benko-2014} and blue points from the K2 mission \citep{garden, moskalik-2018, eap}} \label{fig:5}
\end{figure}

A short cadence RRd target of Campaign 1, EPIC 201585823 has been analysed by \citet{kurtz-2016}. This star provided a new space-based detection of low-amplitude modes in double-mode RR\,Lyraes, in addition to AQ Leo \citep{gruberbauer-2007}, CoRoT ID 0101368812
\citep{chadid-2012} and two more from the K2-E2 data \citep{molnar-2015b}. Although the interpretation of these modes differ in the aforementioned publications, they all belong to the same group of modes with  $P/P_0\sim0.61$ period ratio. \citet{kurtz-2016} also reported the first comparison of photometric pipelines in the case of an RR\,Lyrae, and concluded that the careful choice of photometric mask was essential.
Double-mode RR\,Lyrae stars observed during the K2 Mission have been studied by \citet{moskalik-2018}. 39 RRd stars were investigated from Campaigns 0 to 13 in this preliminary study. The major part of these stars show the typical period ratios, that form a well-defined arc in the Petersen diagram and denoted as 'classical' RRd stars \citep{sos-2019}. Three stars, in turn, have lower period ratios belonging to the newly identified subgroups: two of them to the anomalous RRd stars \citep{sos-2016}
and one to the group identified by \citet{prudil-stars}. The pulsation modes of both anomalous RRd stars, as most of the members of the subgroup, show modulations. Similarly, the detected strong dominance of the fundamental mode is typical for anomalous RRd stars. The 'Prudil' stars are a mysterious group where the mode of shorter period cannot be the radial first overtone. Returning to the classical RRd sample of K2, the $f_{0.61}$ mode has been detected in many of them at the millimagnitude level. Sometimes these modes seem to be non-stationary, while the dominant
radial modes are stable. The additional modes of RRc stars populate two different regions in the Petersen diagram (Fig. \ref{fig:5}.): three sequences of $f_{0.61}$ modes (which according to \cite{dziem-2016} might correspond to non-radial modes
of moderate degree $\ell$ = 8, 9 and the middle sequence due to their linear combination) and the coherent frequency group of $f_{0.68}$ modes recently studied in detail by \citet{netzel-2019}. Theoretical explanation for the latter group is still in question. Only one star have been found so far in which  $f_{0.61}$ and $f_{0.68}$ modes coexist: KIC 9453114 in the \textit{Kepler} field \citep{moskalik-2015}.
Low-amplitude modes of RRab stars from the early Campaigns of the K2 mission have been investigated by \citet{garden}. The Petersen diagram of period ratios of these new findings traced out the new groups of RRab low-amplitude modes. In Fig. \ref{fig:5} we present a Petersen diagram for the various multi-mode RR Lyrae stars of all subtypes, based on the OGLE and the \textit{Kepler}/K2 data. The most unambiguous new group is the $f_2$ group, which shows a much lower scatter in frequencies than the group near $f_1$. Many stars exhibit low-amplitude peaks at slightly longer frequencies than $1.5f_0$ (i. e. above the period doubling line at $\sim$0.666 period ratio in the Petersen diagram), the transition  toward the $f_1$ regime is almost continuous. It is unclear what causes this incredible variety of low-amplitude modes, and the origin of stars outside the main groups is also mysterious. 

As we mentioned, only a small fraction of the K2 RR Lyrae sample is processed so far. Additional frequencies from the whole K2 survey will populate the Petersen diagram even more and may provide us better understanding of the mode selection mechanism.

\begin{figure}[!h]
\begin{center}
\includegraphics[width=175mm]{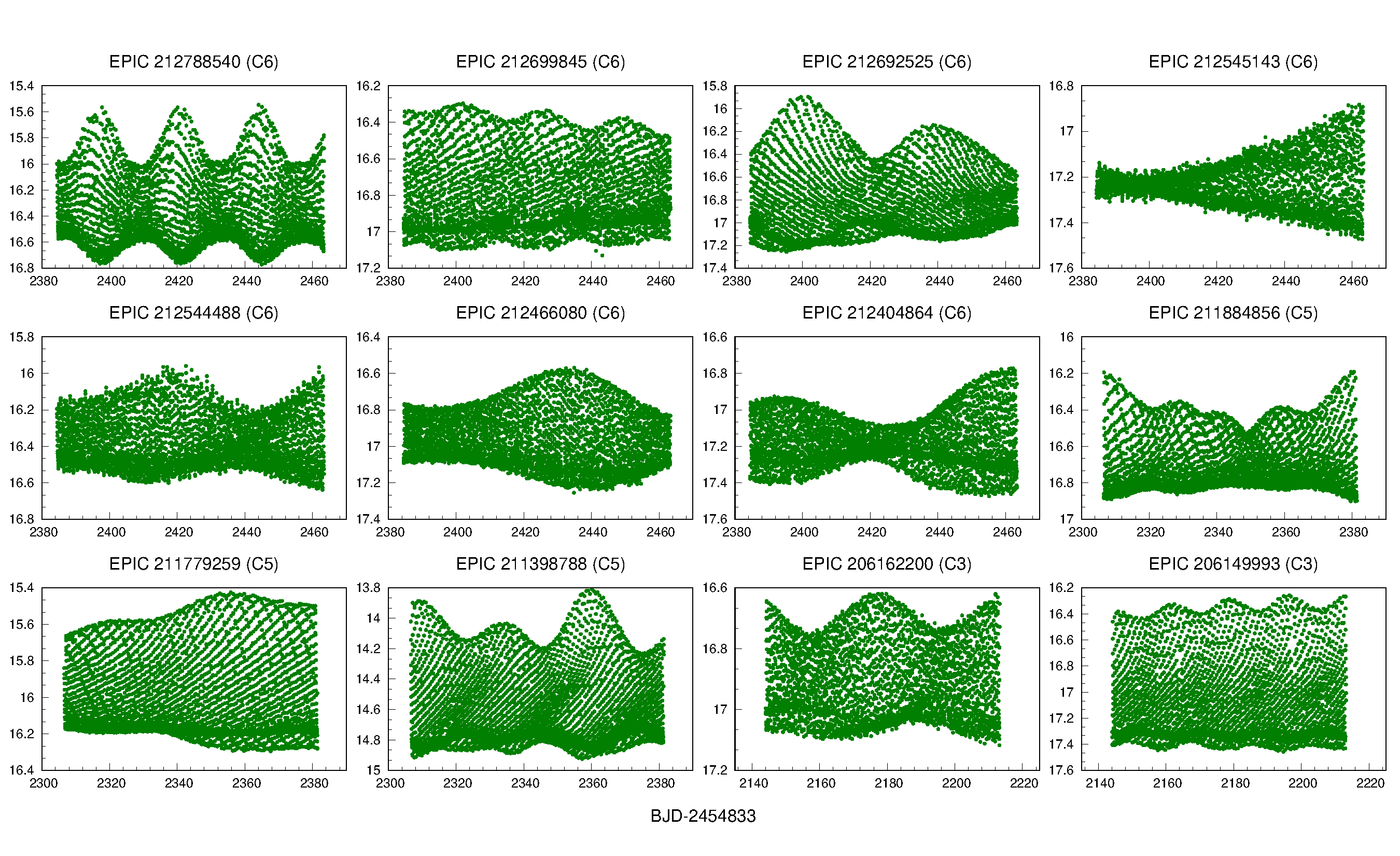}
\end{center}
\caption{Blazhko stars in the K2 mission. Based on the light curves published by \citet{eap}.}\label{fig:6}
\end{figure}

\vspace{15mm}

\subsection{Studying the Blazhko effect with K2}
The high-quality massive photometry of K2 RR\,Lyrae stars offers a basis to investigate the incidence rate of the Blazhko effect. To recover modulation, not only the modulation triplets could be searched in the Fourier spectra, but it is possible to construct proper amplitude and phase variation curves by a template fitting method \citep{eap}. This kind of analysis resulted in 44.7\% for the incidence rate of Blazhko effect that is in agreement with the widely excepted value of around fifty percent. These incidence rates have  been questioned by \citet{kovacs-2018, kovacs-2020} who proposed all RR\,Lyrae stars to be modulated based on his analysis of K2 data produced with the SAP pipeline. However, the estimation of the incidence rate strongly depends on the time span and the data quality, and as we mentioned earlier SAP light curves are of inferior quality compared to EAP light curves, therefore in our opinion it is highly unlikely that a thorough analysis would return close 90\% percent occurrence. The literature values of the RRab Blazhko incidence rate range between 5 to 60 percent (see \cite{kovacs-2016}). The analysis of the most extended sample over 8000 RRab in the Galactic bulge by OGLE IV put the minimum value just above $\sim$40\% \citep{prudil-2017}.

Not only the incidence rate can be studied with K2, but the fine details of the Blazhko effect at the shortest modulation periods as well. The K2 Blazhko sample shows a large variety of modulations, some examples are displayed in Fig. \ref{fig:6}. 

The simultaneous appearance of the Blazhko effect and period doubling in some stars supports the theory of their common origin \citep{buchler-2011}. However, K2 Blazhko stars show a very low fraction of period doubled cases: only 7 stars out of 166 display clear subharmonics in their Fourier spectra. This is a much lower incidence rate that we experienced in the original \textit{Kepler} field (9 out of 15).


\section{Outlook}

After nine years of operation \textit{Kepler} run out of fuel and officially retired on 30 October, 2018 leaving us with an unprecedented amount of space photometric data. Ongoing and upcoming missions, like NASA's original and extended TESS mission \citep{ricker-2015} and the PLATO mission of the European Space Agency  \citep{rauer-2014} scheduled for launch in 2026, will provide hundreds-to-thousands of continuous RR\,Lyrae light curves spanning from a few weeks, to several months to years coverage. 

TESS observes brighter targets than \textit{Kepler} did, partly because of its smaller apertures and larger pixel size (4$^{\prime\prime}$ for \textit{Kepler} and 21$^{\prime\prime}$ for TESS), and its observation length provides shorter (27 days) light curves than K2 did, for most of the targets. But even with these short observations $\sim$30--130 RR~Lyrae pulsation cycles can be monitored, adequate to explore the low-amplitude mode content. One year of quasi-continuous coverage is also possible close to the ecliptic poles, lending  opportunity to study the Blazhko effect in this regions. In addition, large parts of the sky will be re-observed by TESS during its extended mission(s), hence altogether 90\% of the sky will be covered.
 PLATO will also cover much larger areas of the sky than the  \textit{Kepler} and K2 missions: it will have long observing runs (up to 1-2 years) complemented by shorter (2-3 months) so-called step-and-stare runs, thus altogether nearly 50\% of the sky will be covered.

Similarly,  Gaia \citep{gaia-2018}, and the Legacy Survey of Space and Time (LSST) of the Vera Rubin Observatory \citep{lsst-2009} are and will be game-changers in the field. Gaia provides parallax and proper motion information for more than 1.3 billion stars down to 21 magnitude complemented by variability information, basic color measurements and low-resolution spectroscopy in Data Release 2, and the numbers will increase in the later releases. Tens of thousands of new RR Lyrae stars will be discovered providing a homogeneous all-sky catalog which is not possible by either of the aforementioned missions. Coincidentally, Gaia will provide distances down to the faint end of TESS's RR Lyrae sample. LSST will monitor the sky visible form Chile in six photometric bands every 3 days with a 3.5 Gpixel camera for 10 years, thus providing on average 800–1000 photometric data points (i.e. OGLE-like cadence) for each targets down to 23.5 mag (reaching 25 mag in co-added images).
This capability will allow the discovery of tens of thousands of new RR Lyrae stars out to a large distance (400 kpc) providing excellent opportunities to use RR Lyrae stars to trace halo structures in the galactic halos and discover faint surface brightness dwarf galaxies in the Local Group. 

All these prospective data sets from upcoming missions will shed new light on the occurrence of the Blazhko effect (including long-period and multi-period modulations), period doubling, additional radial and nonradial modes, chaotic behavior, and other dynamical phenomena as a function of a broad range of stellar parameters, Galactic and extragalactic environments. In light of these prospects, we are witnessing a golden era of classical pulsating variables, including RR\,Lyrae stars.

\section{Funding}
This project has been supported by the Lend\"ulet Program  of the Hungarian Academy of Sciences, projects No. LP2014-17 and LP2018-7/2020 and the MW-Gaia COST Action (CA-18104). This paper includes data collected by the \textit{Kepler} mission and obtained from the MAST data archive at the Space Telescope Science Institute (STScI). EP acknowledges the financial support of the Hungarian National Research, Development and Innovation Office (NKFI), grant KH\_18 130405 and the J\'anos Bolyai Research Scholarship of the Hungarian Academy of Sciences. This research has made use of NASA’s Astrophysics Data System. Funding for the \textit{Kepler} mission is provided by the NASA Science Mission Directorate. STScI is operated by the Association of Universities for Research in Astronomy, Inc., under NASA contract NAS 5–26555. 

\section{Acknowledgments}
The authors thank the reviewers for their valuable comments and suggestions that helped to improve this paper. The authors gratefully acknowledge the entire \textit{Kepler} team, whose outstanding efforts have made these results possible. Fruitful discussions with L\'aszl\'o Moln\'ar and Zolt\'an Koll\'ath are also acknowledged.
\bibliographystyle{frontiersinSCNS_ENG_HUMS} 
\bibliography{main}
\end{document}